# Superconducting transition of FeSe/SrTiO$_3$ induced by adsorption of semiconducting organic molecules

Jiaqi Guan,[1] Jian Liu,[1] Bing Liu,[1] Xiaochun Huang,[1] Qing Zhu,[1] Xuetao Zhu,[1] Jiatao Sun,[1] Sheng Meng,[1,2] Weihua Wang,[1]* and Jiandong Guo[1,2,3]*

[1]*Beijing National Laboratory for Condensed Matter Physics and Institute of Physics, Chinese Academy of Sciences, Beijing 100190, China*
[2]*Collaborative Innovation Center of Quantum Matter, Beijing 100871, China*
[3]*School of Physical Sciences, University of Chinese Academy of Sciences, Beijing 100190, China*

*Email: weihuawang@iphy.ac.cn and jdguo@ iphy.ac.cn

**ABSTRACT**

We prepared superconducting and non-superconducting FeSe films on SrTiO$_3$(001) substrates (FeSe/STO) and investigated the superconducting transition induced by charge transfer between organic molecules and FeSe layers by low temperature scanning tunneling microscopy and spectroscopy. At low coverage, donor- and acceptor-type molecules adsorbed preferentially on the non-superconducting and superconducting FeSe layers, respectively. Superconductivity was induced by donor molecules on non-superconducting FeSe layer, while the superconductivity was suppressed near acceptor molecules. The corresponding evolutions of electronic states and work function were also resolved by scanning tunneling microscopy. These results illustrate the important role played by local electron concentration in the superconducting transition of FeSe/STO.

# I. INTRODUCTION

The enhanced superconductivity of single-layer FeSe grown on $SrTiO_3(001)$ substrates (FeSe/STO) has attracted intense research interests both in experiment and theory [1-12]. The electron concentration in FeSe layers is believed to play a vital role in the superconducting transition of FeSe/STO [5-8], and tremendous efforts have been devoted to tune the electron concentration by post-annealing [5, 7, 13, 14], alkali metal atom deposition [15-18] and gate voltage [8, 14, 19]. Organic molecules adsorption has been widely used to tune the charge carrier densities in transition metal chalcogenides [20], graphene [21, 22], topological insulators [23] and cuprate superconductor films [24]. The electron concentration in substrates were either increased or decreased by depositing donor- or acceptor-type molecules [22, 25].

Organic molecules may also provide a much convenient way to tune the electron concentration in FeSe/STO. Compared to alkali metal atoms, organic molecules are stable in ambient conditions, and the substrate does not need to be held at low temperature during deposition. Furthermore, the structural effect of FeSe/STO can be minimized by choosing closed-shell molecules physisorbed on the substrate, *i.e.* via van der Waals interaction, enabling us to study the superconducting transition primarily driven by local electron concentration. On the other hand, the post-annealed FeSe/STO substrate, which has co-existing electron-doped 1st layer and nearly neutral 2nd layer [5-7, 26], may introduce interesting adsorption phenomena to donor or acceptor molecules.

Here we report the selective adsorption behaviors of typical donor- and acceptor-type molecules on ultra-thin FeSe/STO with co-existing superconducting (SC) and non-superconducting (non-SC) areas. Local work function (LWF) measurements and density functional theory (DFT) calculations reveal the interfacial charge transfer induced by the donor- and acceptor-type molecules. The corresponding superconductivity inducement/suppression due to the increase/decrease of local electron density is observed. Additionally, DFT calculations reveal the interfacial charge transfer without significant lattice modification, suggesting that those organic molecules are good candidates for electronic tuning of superconductivity on FeSe/STO.

## II. EXPERIMENTAL AND THEORETICAL METHODS

The experiments were performed in a Unisoku ultra-high vacuum (UHV) low temperature scanning tunneling microscope system combined with a molecular beam epitaxy (MBE). The FeSe films were grown on Nb-doped (0.5% wt) $SrTiO_3$(001) (STO) substrates by the reported method [1]. The as-grown samples were post-annealed at 470 ºC for 6 h in UHV condition to make the first layer FeSe superconducting [1, 5, 26]. We used typical donor and acceptor molecules dibenzotetrathiafulvalene (DBTTF) and 7,7,8,8-tetracyanoquinodimethane (TCNQ) [22], whose chemical structures are depicted in Fig. 1(a). DBTTF and TCNQ molecules were evaporated onto the as-grown or annealed FeSe/STO sample from evaporators at 410 K and 390 K, respectively. After deposition, the sample was transferred into the cryostat of STM with a base pressure better than $1.0 \times 10^{-10}$ Torr. Polycrystalline Pt-Ir tips, cleaned by electron beam bombardment and verified on Ag/Si(111), were used in the scanning tunneling microscopy and spectroscopy (STM/STS) measurements. The STM topographic images were acquired in constant-current mode with the bias voltage applied to the sample with respect to the tip. Unless otherwise specified, the experiments were performed at 4.9 K and the STS were measured with a bias modulation of 1 mV at 987.5 Hz.

First-principle calculations were performed using the VASP code [27-29]. The interactions between valence electrons and ionic cores were described with the projector augmented wave (PAW) method [30]. We adopted the generalized gradient approximation (GGA) with the Perdew-Burke-Ernzerhof (PBE) formula for the exchange-correlation functional [31]. The electron wave functions were expanded in plane wave basis with an energy cutoff of 400 eV. The molecule/FeSe system was modeled using a slab model containing an isolated molecule adsorbed on a single-layer FeSe. The supercell of FeSe layer is $6 \times 4 \times 1$ and $4 \times 4 \times 1$ for DBTTF and TCNQ respectively. A vacuum layer of thickness more than 15 Å was used. All molecule/FeSe structures were fully optimized by allowing all degrees of freedom of the systems to relax until the force acting on each atom was smaller than 0.05 eV/Å. The tetrahedron method with Blöchl corrections [32] was used in the total energy calculations to achieve a high accuracy. In view of the larger computational expense, only the Γ point sampling in the Brillouin zone was performed.

## III. RESULTS AND DISCUSSION

### A. Selective adsorption

We prepare 4 types of FeSe/STO samples for comparison – annealed single-layer FeSe/STO that is superconducting (SC) uniformly, as-grown single-layer FeSe/STO that is non-superconducting (non-SC) uniformly, annealed FeSe/STO with 2nd layer FeSe islands on the first layer that contains SC (1st layer) and non-SC areas (2nd layer), and as-grown FeSe/STO with 2nd layer FeSe islands on the first layer that is non-SC uniformly but with distinct topography on the surface (see Fig. S1 in Supplemental Material). As an example, Fig. 1(b) shows the STM image of a post-annealed FeSe/STO sample with coexisting 1st and 2nd FeSe layers. After post-annealing, the 1st layer shows the characteristic of trench-like defect lines, which distinguishes from the 2nd layer distributed along the step edge, nearly intact except for few isolated Se vacancies on it [1, 2, 13]. As shown in Fig. S2 of Supplemental Material, the STS measured on the 1st layer show a superconducting gap of ~ 20 meV, while the 2nd layer exhibits a non-superconducting semiconductor or bad metal-like electronic structure near the Fermi level, in agreement with previous reports [1].

When the donor- or acceptor-type molecules are deposited on the FeSe/STO samples, selective adsorption behaviors are clearly observed – donor-type DBTTF molecules adsorb on non-SC area preferentially while acceptor-type TCNQ molecules adsorb on SC area preferentially. More specifically, on post-annealed sample, DBTTF molecules adsorb dominantly on the non-SC 2nd layer and form islands at a low molecular coverage [Fig. 1(c)], while the TCNQ molecules adsorb preferentially on the SC 1st FeSe layer of the post-annealed sample [Fig.1(d)]. But on the non-SC as-grown sample, DBTTF molecules adsorb on both the 1st and 2nd layer, as shown in the inset of Fig. 1(c). These organic molecules are closely packed, forming self-assembled DBTTF islands or random TCNQ clusters on FeSe/STO, indicating the weaker molecule-substrate interaction than inter-molecule interaction. Detailed descriptions are presented in Supplemental Material.

The selective adsorption behaviors of DBTTF and TCNQ molecules are essentially related to the local electron density of the tested area, *i.e.*, whether the area is SC or not, rather than the morphology or thickness of the FeSe films. This can be interpreted by the charge-transfer property of the molecules and the electron concentration in respective FeSe regions. The donor-type DBTTF molecules tend to donate electrons to the substrate, and consequently are preferentially adsorbed on the non-SC area where the local electron concentration is lower than the SC area [5-7, 26]. In contrast, the acceptor-type TCNQ molecules tend to extract electrons from the substrate, and are preferentially adsorbed on the SC area where the electron concentration is higher. Similar selective adsorption behaviors induced by charge density inhomogeneity have been reported – an acceptor molecule, $F_{16}CuPc$, preferred to adsorb on more negatively charged monolayer graphene than bilayer graphene on SiC substrate [33, 34].

**B Charge Transfer between molecules and FeSe/STO**

To reveal the electronic modification induced by molecules adsorption, we firstly measure STS curves in a large bias range, as shown in Fig. 2(a). The non-SC pristine 2nd layer FeSe shows a prominent electronic state at -0.15 V. When tip is laterally moved towards the DBTTF island, this feature gradually fades out. And finally the STS curve becomes similar to that measured on the SC 1st layer.

The electron doping from DBTTF molecules to FeSe is evidenced by the local work function (LWF) measurements. The averaged work function is extracted by fitting the tunneling conductance $G$ ($G=I/V$) by $G \propto \exp(-2\sqrt{2m\Phi}z/\hbar)$, where $m$ is the electron mass, and $\Phi=(\Phi_{tip}+\Phi_{sample})/2$ is the averaged work function of tip and sample. Since the work functions are measured with the same tip, the obtained averaged work function is used to evaluate the LWF of respective sites. Figure 3(b) shows the typical $G$-$z$ curves measured at 0.5 V on the SC 1st layer, bare non-SC 2nd layer and the region near the DBTTF island. By repeated measurements on different sites, the averaged work functions of those areas are determined to be 4.91 ± 0.05 eV, 5.11 ± 0.08 eV and 4.98 ± 0.10 eV, respectively. The lowered LWF near the DBTTF molecule is attributed to the lift of Fermi level by electron doping from DBTTF molecules. Similarly, it can be deduced that the lower work function on the 1st layer than the 2nd layer is due to the electron doping from the STO substrate, which is in agreement with the conclusion of previous ARPES research [6, 7].

DFT calculations are performed to further investigate the charge transfer between DBTTF/TCNQ molecules and the single-layered FeSe. Based on total energy calculation, the optimized adsorption configurations for DBTTF/FeSe and TCNQ/FeSe are shown in Fig. 3(a) and (b), respectively. The two molecules are both physically adsorbed on FeSe via weak van der Waals interaction, and the maximum distance between the molecular plane and top Se atoms in FeSe is 3.72 Å for DBTTF/FeSe and 3.41 Å for TCNQ/FeSe. The physisorption of both molecules are in agreement with experimental observations (see Supplemental Material).

To study the charge redistribution upon the molecular adsorption on FeSe substrate, differential charge density (DCD) of the most stable adsorption configuration for each molecule/FeSe system is calculated by [22]

$$\Delta \rho_{mol/FeSe} = \rho_{mol/FeSe} - \rho_{mol} - \rho_{FeSe}, \qquad (1)$$

where $\rho_{mol/FeSe}$, $\rho_{mol}$, and $\rho_{FeSe}$ are the charge density of the adsorbed system, the charge density of the molecule, and the charge density of the single-layer FeSe, respectively. The results are also shown in Figs. 3(a) and 3(b). The charge accumulation region mainly locates around DBTTF for DBTTF/FeSe, while the charge depletion region locates around the TCNQ molecule for TCNQ/FeSe, clearly indicating the opposite charge transfer directions of the two molecule/FeSe systems. We define the planar averaged DCD as $\Delta\rho_{mol/FeSe}(z) = \int \Delta\rho_{mol/FeSe}(x,y,z)dxdy$ [see the calculation results in Fig. 3(c) and (d)]. Integrating DCD via $Q = \int_z^0 \Delta\rho_{mol/FeSe}(z)\,dz$, we obtain the amount of the interfacial charge transfer quantitatively, *i.e.*, the amount of the interfacial charge transfer from one molecule to the substrate for DBTTF/FeSe and TCNQ/FeSe are 0.095 $e$ and -0.39 $e$, respectively.

The above planar averaged DCD method neglects the differential charge density between molecule and FeSe. We further carry out the Bader charge analysis to estimate the interfacial charge transfer [35]. The results show that for the DBTTF/FeSe system, the FeSe substrate obtains 0.05 $e$ and one DBTTF molecule loses 0.24 $e$ comparing to their neutral state, with 0.19 $e$ delocalized around the interface between them. For the TCNQ/FeSe system, the FeSe substrate loses 0.54 $e$ and one TCNQ molecule obtains 0.34 $e$. Previous ARPES results showed that ~0.1 $e$ per FeSe unit cell has been transferred from STO substrate via sufficiently annealing [5, 7]. Considering the larger size of the DBTTF molecules (12.6 nm × 0.5 nm per molecule) than a FeSe unit cell, the donor-type molecule has a weaker capability than STO for electron doping to FeSe.

## C Superconducting transition induced by molecules

Considering the van der Waals interaction and charge transfer between the molecules and FeSe layer, the DBTTF and TCNQ molecules on FeSe/STO provides an arena to study the influence of electron concentration on superconducting transition without significant effect on the local lattice structure. Figure 4(a) shows a zoom-in image of DBTTF island on the 2nd layer FeSe of a post-annealed sample. The FeSe lattice can be clearly recognized in the image, as the unit vectors being indicated by the arrows in the lower left corner. A typical differential conductance d$I$/d$V$ curve near the DBTTF island (within the distance of 2 $a$, $a$ is the FeSe lattice parameter) is plotted in the upper panel of Fig. 4(b), and the normalized curve is plotted in the bottom panel, which was derived following the method described in Ref.

[16]. Two pronounced peaks located at ±8 meV and the zero conductance with zero-bias clearly indicate that the superconductivity is induced by electron doping, although the superconducting gap is reduced relative to that measured on the annealed 1st layer FeSe. The normalized d$I$/d$V$ spectra taken on the 2nd layer ~1$a$ from DBTTF island at elevated temperatures from 4.9 K to 24.6 K are plotted in Fig. 3(c). With temperature increasing, the two coherence peaks become weak and gradually merge with each other, while the zero-bias conductance increases. Thus the superconducting transition temperature is determined to be around 25 K, which is remarkably higher than the bulk FeSe [36, 37].

In agreement with the differential conductance spectra in large bias range, LWF measurement, as well as DFT calculations that reveal the electron accumulation around DBTTF molecules, the superconducting transition induced by electron doping from DBTTF shows a clear site dependence. As shown in Fig. 4(d) (see Fig. S5 in Supplemental Material for the all the STS curves measured along the arrow AB), when the tip is positioned 15 $a$ away from the DBTTF island, the STS curve shows a semiconducting feature with two electronic states at 11 mV and 26 mV, and the occupied density of states are less pronounced than the unoccupied states in the range of -50 mV to 50 mV. When the distance between tip and DBTTF molecules is less than 3 $a$, superconducting gaps are observed in STS, manifested by two peaks distributed symmetrically with respect to the Fermi level and the zero conductance with zero-bias. These results indicate the coexistence of the non-SC (pristine 2nd layer FeSe) and SC (around the DBTTF molecules) areas on the surface.

On as-grown FeSe/STO, the superconductivity can also be induced by DBTTF molecules. Figure 4(e) shows the STM image of DBTTF molecules adsorbed on the as-grown 1st layer FeSe. The STS measured on bare 1st layer shows a depression of electron states at Fermi level without any sign of coherence peaks [Fig. 4(f)], indicating the non-superconducting characteristics. With tip approaching to the DBTTF island within 3 $a$, two coherence peaks and a U-shaped gap emerge in STS curves. The STS curve measured at 1 $a$ from DBTTF molecules shows a superconducting gap of ~ 10 meV, indicating the superconducting transition being induced on as-grown FeSe/STO.

On the other hand, the superconductivity of FeSe/STO is also influenced by the adsorption of acceptor-type molecule TCNQ. Figure 5(a) is an STM image of TCNQ molecules on the SC 1st layer FeSe. At the point about 11 *a* away from the TCNQ molecules, the STS shows a U-shaped superconducting gap at the Fermi level and two coherence peaks at ±15 meV. As the tip laterally approaching the TCNQ molecules, the superconductivity become suppressed – the gap becomes asymmetric with respect to the Fermi level, and the coherence peak above the Fermi level gradually vanishes, until semiconducting characteristics appears. On properly annealed FeSe/STO, which shows enhanced the superconductivity, a superconductor-semiconductor transition can be induced by decreasing the local electron concentration.

**D Discussions**

Different from K atoms, which tend to adsorb individually at low coverage and spread all over the substrate [16-18], DBTTF (TCNQ) molecules coalesce into islands (clusters) and leave bare FeSe areas, which enable us to study a localized superconducting transition with spatial resolution. The superconductivity on the 1st layer FeSe of as-grown sample induced by DBTTF molecules also highlights the difference between DBTTF and K atoms in electron doping of FeSe/STO – the attempts to induce superconductivity on non-superconducting first layer FeSe have been failed so far [17]. This may probably stem from the different influence on FeSe lattice by K doping and molecule adsorption.

It should be noted that the DBTTF molecule has much lower charge transfer ability than the K atoms. Our DFT calculation reveals that the FeSe layer gets 0.095 e from one DBTTF molecule based on averaged DCD calculation, or the FeSe layer gets 0.05 e and the DBTTF molecule loses 0.24 e by Bader charge analysis. In contrast, it can be estimated that one K atom loses one electron to the substrate [16]. Although the electron doping ability of DBTTF molecules are much lower than that of K atoms, DBTTF molecules are stable in ambient conditions and can be easily removed from FeSe substrate by moderate annealing. Therefore they are promising candidates as capping layers to protect FeSe/STO not only in *ex situ* transport measurements, but also in sample transfer between different vacuum systems instead of Se that suppresses the superconductivity.

## IV. CONCLUSIONS

In summary, the superconductivity of FeSe/STO is tuned on/off by adsorption of donor-type DBTTF or acceptor-type TCNQ molecules, which modifies the local electron concentration. Besides of selective adsorption behaviors of the molecules, our work emphasizes the role of local electron concentration in the superconductivity enhancement in FeSe/STO. Technically, DBTTF molecules are promising candidates to be used as capping layers to protect FeSe/STO in *ex situ* measurements.

**Acknowledgements**

This work is supported by the Hundred Talents Program of the Chinese Academy of Sciences. J. G. is grateful to the financial support of Chinese NSFC (11634016 & 11474334) and the "Strategic Priority Research Program (B)" of the Chinese Academy of Sciences (XDB07030100).

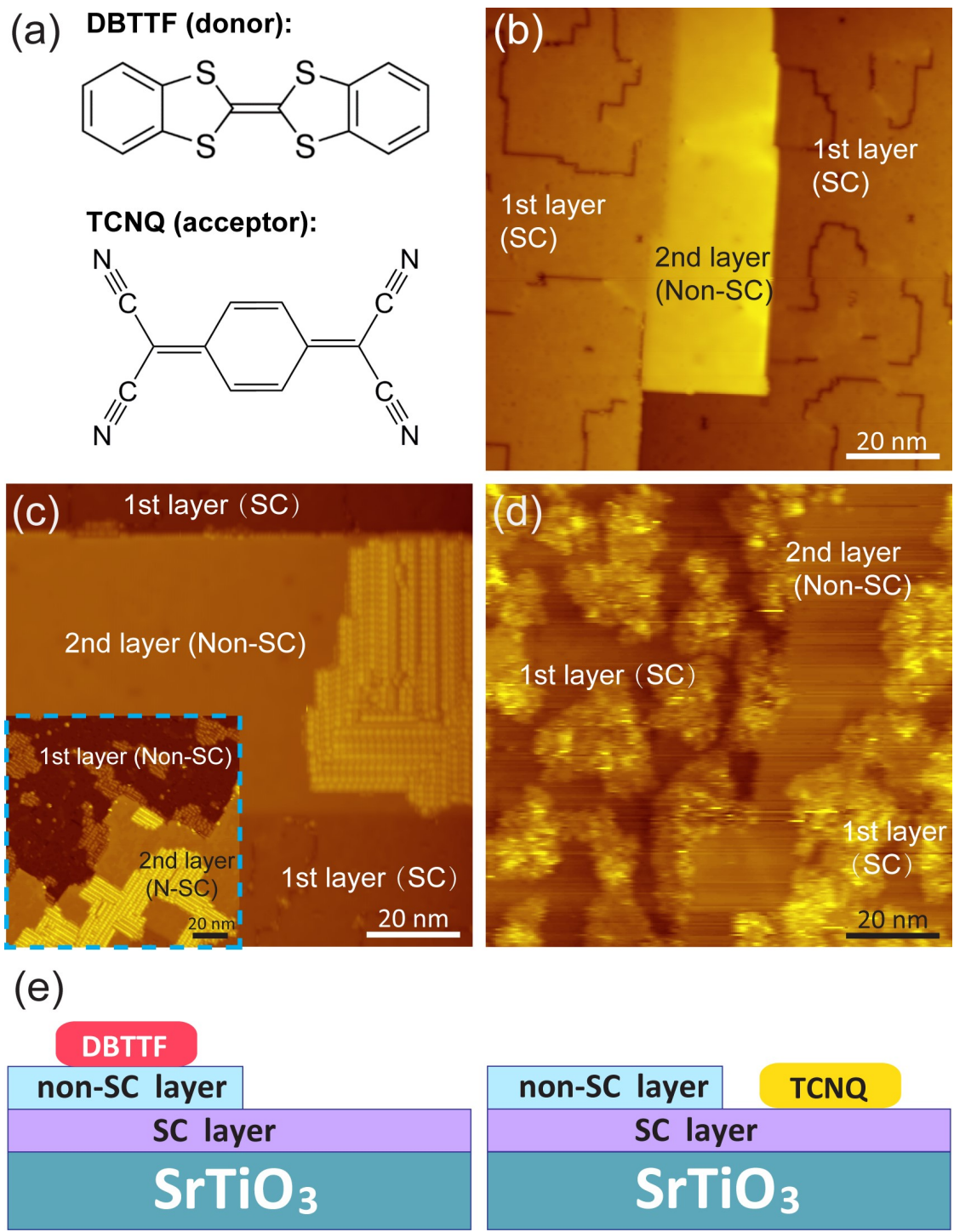

FIG. 1. (Color online). (a) Chemical structures of DBTTF and TCNQ molecule. (b) STM image (2.0 V / 50 pA) of a post-annealed FeSe/STO sample. (c) STM image (2.5 V / 30 pA) of DBTTF molecules on post-annealed FeSe/STO. Inset: STM image (2.0 V / 50 pA) of DBTTF molecules on as-grown FeSe/STO. (d) STM image (5.0 V / 50 pA) of TCNQ molecules on post-annealed FeSe/STO. (e) Adsorption schematics of the DBTTF/TCNQ molecules on the non-SC (2nd layer) / SC (1st layer) post-annealed FeSe/STO.

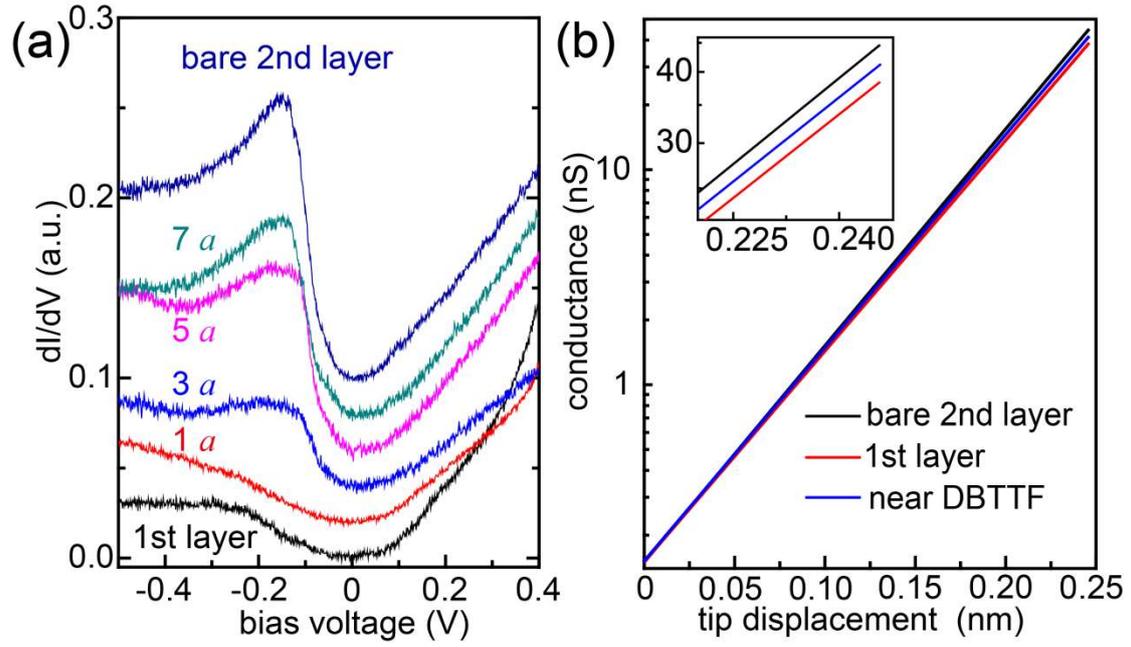

FIG. 2. (Color online) (a) Averaged d$I$/d$V$ spectra in a large bias range measured on the 2nd layer FeSe near a DBTTF island. A spectrum measured on the 1st layer FeSe is shown for comparison. The curves are vertically shifted. $a$ is the FeSe lattice parameter. (b) Typical $G$-$z$ curves measured on the superconducting 1st layer, bare non-superconducting 2nd layer and the region near DBTTF, with a zoom-in part in the inset. All these curves were measured with a fixed bias voltage of 0.5 V.

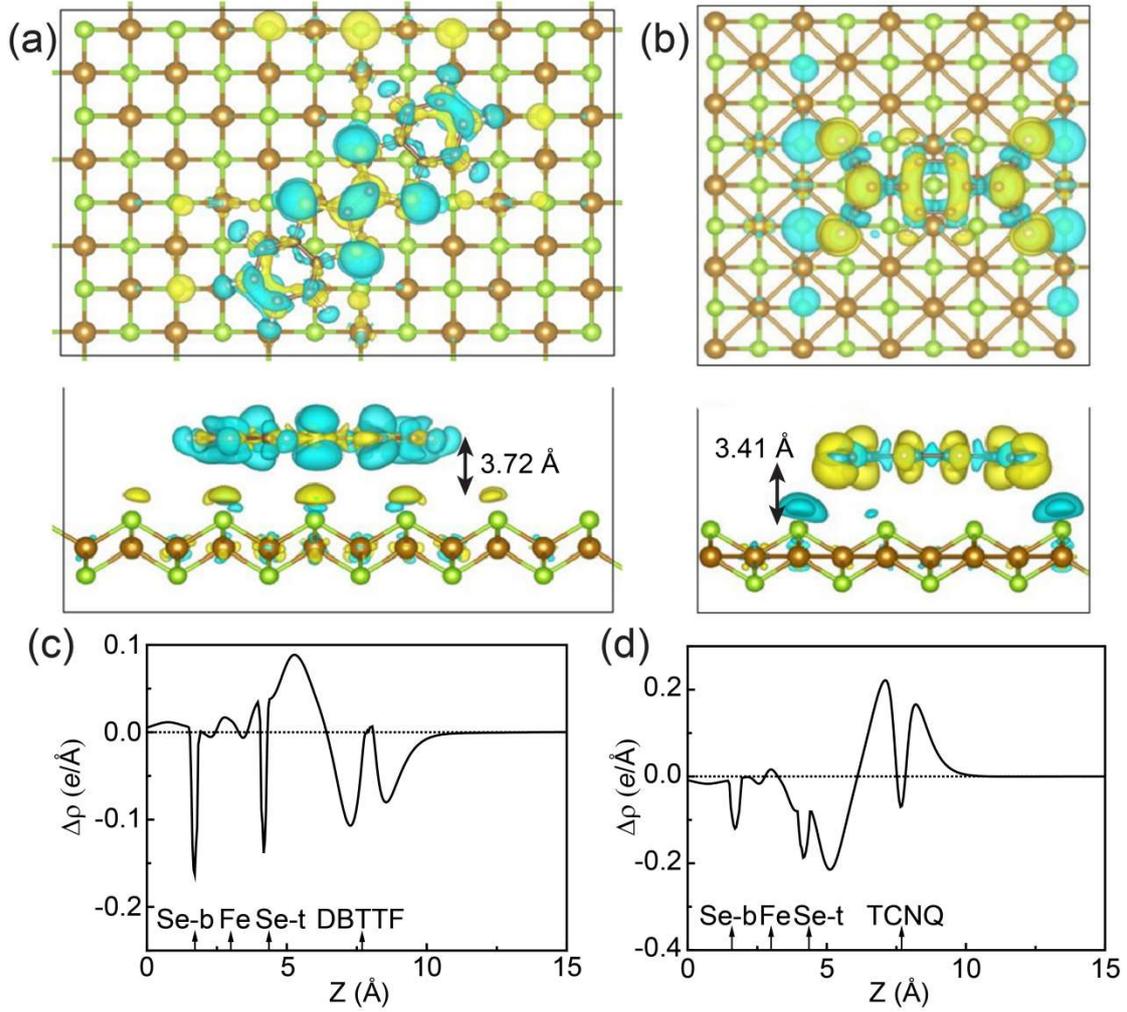

FIG. 3. (Color online) The top views (upper) and side views (bottom) of the charge rearrangement of (a) DBTTF and (b) TCNQ on single-layer FeSe. Cyan (yellow) color indicates charge depletion (accumulation) region. (c) and (d) show the planar averaged charge density $\Delta\rho$ induced by DBTTF and TCNQ adsorption on single-layer FeSe, respectively. The arrows on the bottom axis indicate the positions of bottom Se atoms (Se-b), Fe atoms, top Se atoms (Se-t), and DBTTF/TCNQ molecules in z direction, respectively.

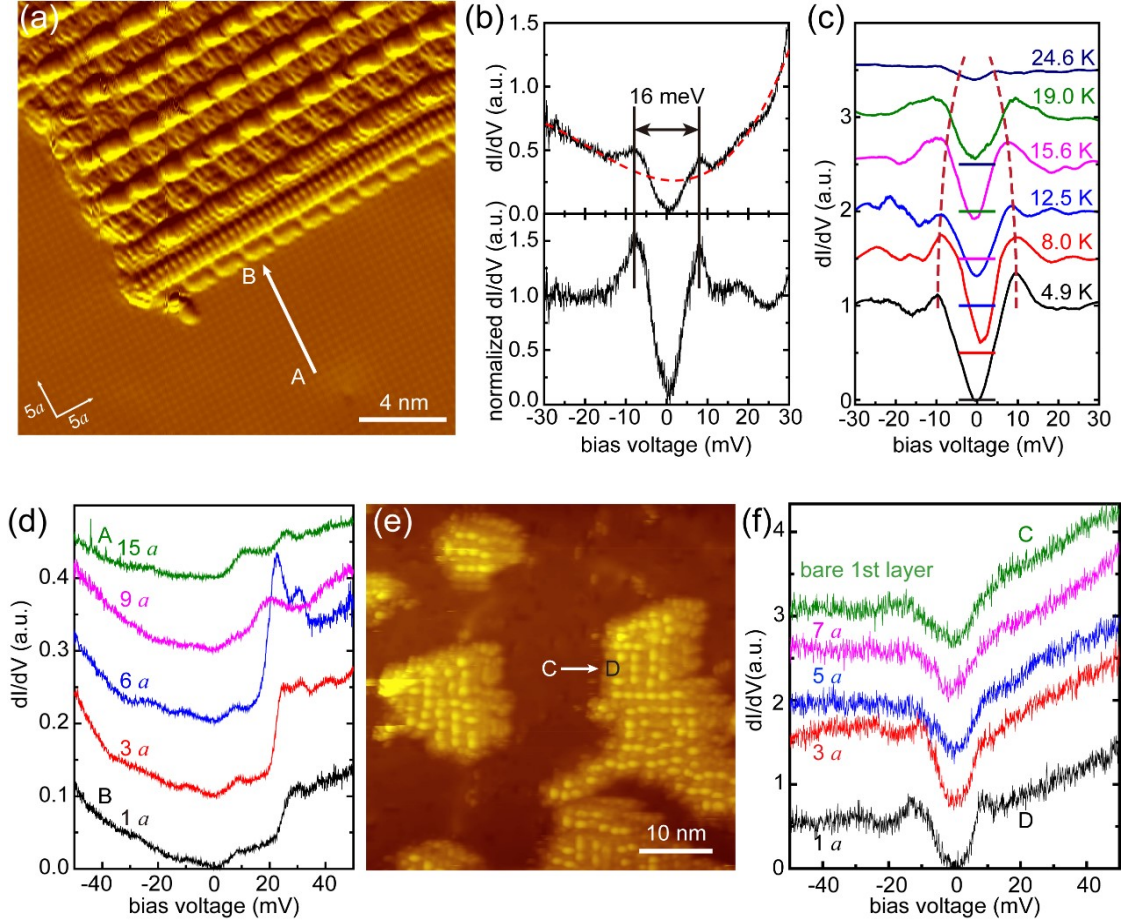

FIG. 4. (Color online) (a) STM image (20 × 20 nm$^2$, 1.5 V / 50 pA, differentiated) of DBTTF molecules adsorbed on the second layer FeSe. (b) The STS shows the induced superconductivity near DBTTF molecules on the non-SC 2nd layer FeSe. The upper panel plots the raw data, while the lower panel plots the normalized data following the method described in Ref. [16]. The background for normalization is plotted in red dashed line. (c) Normalized d$I$/d$V$ spectra taken on the 2nd layer FeSe ~1 $a$ from the DBTTF island at varied temperatures. The dash lines show the synchronous change of the coherence peaks. (d) d$I$/d$V$ spectra taken along AB in (a). The curves are vertically shifted for presentation. (e) STM image (50 × 50 nm$^2$, 2.5 V / 50 pA) of DBTTF molecules adsorbed on the as-grown 1st layer FeSe. (f) d$I$/d$V$ spectra taken along arrow CD in (e). The curves are vertically shifted.

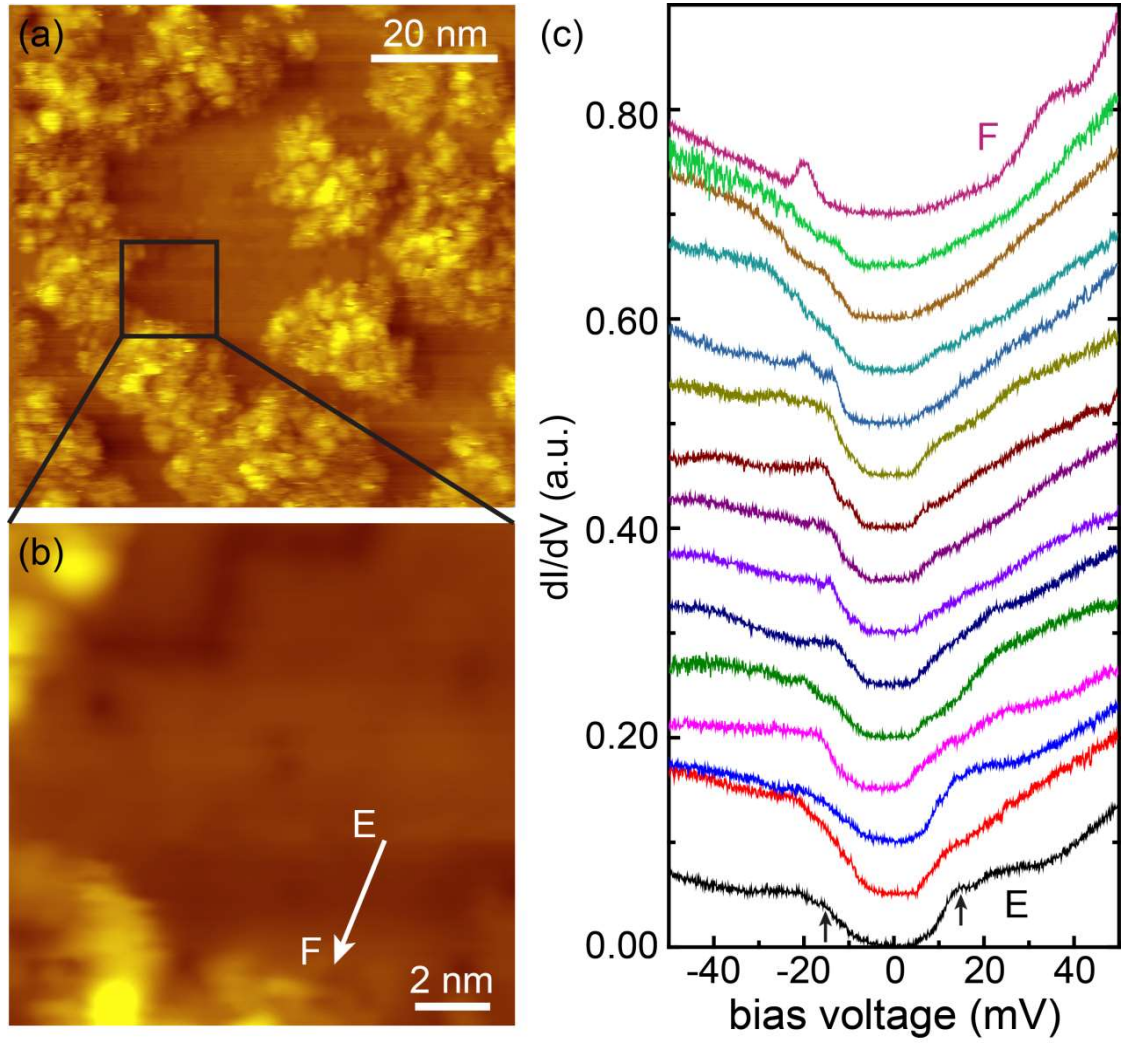

FIG. 5. (Color online) (a) STM image (80 × 80 nm$^2$, 5.0 V / 50 pA) of TCNQ molecules adsorbed on the 1st layer FeSe. (b) Zoom-in image (15 × 15 nm$^2$, 2.0 V / 50 pA) of the square in (a). (c) d$I$/d$V$ spectra taken along EF in (b). The black arrows indicate the positions of coherence peaks at E. The curves are vertically shifted.

**REFERENCE**

[1] Q.-Y. Wang, Z. Li, W.-H. Zhang, Z.-C. Zhang, J.-S. Zhang, W. Li, H. Ding, Y.-B. Ou, P. Deng, K. Chang, J. Wen, C.-L. Song, K. He, J.-F. Jia, S.-H. Ji, Y.-Y. Wang, L.-L. Wang, X. Chen, X.-C. Ma, and Q.-K. Xue, Chin. Phys. Lett. **29**, 037402 (2012).

[2] W.-H. Zhang, Y. Sun, J.-S. Zhang, F.-S. Li, M.-H. Guo, Y.-F. Zhao, H.-M. Zhang, J.-P. Peng, Y. Xing, H.-C. Wang, T. Fujita, A. Hirata, Z. Li, H. Ding, C.-J. Tang, M. Wang, Q.-Y. Wang, K. He, S.-H. Ji, X. Chen, J.-F. Wang, Z.-C. Xia, L. Li, Y.-Y. Wang, J. Wang, L.-L. Wang, M.-W. Chen, Q.-K. Xue, and X.-C. Ma, Chin. Phys. Lett. **31**, 017401 (2014).

[3] D. Liu, W. Zhang, D. Mou, J. He, Y.-B. Ou, Q.-Y. Wang, Z. Li, L. Wang, L. Zhao, S. He, Y. Peng, X. Liu, C. Chen, L. Yu, G. Liu, X. Dong, J. Zhang, C. Chen, Z. Xu, J. Hu, X. Chen, X. Ma, Q. Xue, and X. J. Zhou, Nat. Comm. **3**, 931 (2012).

[4] K. Liu, Z.-Y. Lu, and T. Xiang, Phys. Rev. B **85**, 235123 (2012).

[5] S. He, J. He, W. Zhang, L. Zhao, D. Liu, X. Liu, D. Mou, Y.-B. Ou, Q.-Y. Wang, Z. Li, L. Wang, Y. Peng, Y. Liu, C. Chen, L. Yu, G. Liu, X. Dong, J. Zhang, C. Chen, Z. Xu, X. Chen, X. Ma, Q. Xue, and X. J. Zhou, Nat. Mater. **12**, 605 (2013).

[6] S. Tan, Y. Zhang, M. Xia, Z. Ye, F. Chen, X. Xie, R. Peng, D. Xu, Q. Fan, H. Xu, J. Jiang, T. Zhang, X. Lai, T. Xiang, J. Hu, B. Xie, and D. Feng, Nat. Mater. **12**, 634 (2013).

[7] J. He, X. Liu, W. Zhang, L. Zhao, D. Liu, S. He, D. Mou, F. Li, C. Tang, Z. Li, L. Wang, Y. Peng, Y. Liu, C. Chen, L. Yu, G. Liu, X. Dong, J. Zhang, C. Chen, Z. Xu, X. Chen, X. Ma, Q. Xue, and X. J. Zhou, Proc. Nat. Acad. Sci. USA **111**, 18501 (2014).

[8] W. Zhao, M. Li, C.-Z. Chang, J. Jiang, L. Wu, C. Liu, Y. Zhu, J. S. Moodera, and M. H. W. Chan, arXiv **1701.03678** (2016).

[9] J. J. Lee, F. T. Schmitt, R. G. Moore, S. Johnston, Y.-T. Cui, W. Li, M. Yi, Z. K. Liu, M. Hashimoto, Y. Zhang, D. H. Lu, T. P. Devereaux, D.-H. Lee, and Z.-X. Shen, Nature **515**, 245 (2014).

[10] D.-H. Lee, Chin. Phys. B **24**, 117405 (2015).

[11] S. Zhang, J. Guan, X. Jia, B. Liu, W. Wang, F. Li, L. Wang, X. Ma, Q. Xue, J. Zhang, E. W. Plummer, X. Zhu, and J. Guo, Phys. Rev. B **94**, 08116(R) (2016).

[12] Z.-X. Li, F. Wang, H. Yao, and D.-H. Lee, Sci. Bull. **61**, 925 (2016).

[13] Z. Li, J.-P. Peng, H.-M. Zhang, W.-H. Zhang, H. Ding, P. Deng, K. Chang, C.-L. Song, S.-H. Ji, L. Wang, K. He, X. Chen, Q.-K. Xue, and X.-C. Ma, J. Phys.: Condens. Matter **26**, 265002 (2014).

[14] W. Zhang, Z. Li, F. Li, H. Zhang, J. Peng, C. Tang, Q. Wang, K. He, X. Chen, L. Wang, X. Ma, and Q.-K. Xue, Phys. Rev. B **89**, 060506(R) (2014).

[15] Y. Miyata, K. Nakayama, K. Sugawara, T. Sato, and T. Takahashi, Nat. Mater. **14**, 775 (2015).

[16] C. Tang, C. Liu, G. Zhou, F. Li, H. DIng, Z. Li, D. Zhang, Z. Li, C. Song, S. Ji, K. He, L. Wang, X. Ma, and Q.-K. Xue, Phys. Rev. B **93**, 020507(R) (2015).

[17] C. Tang, D. Zhang, Y. Zang, C. Liu, G. Zhou, Z. Li, C. Zheng, X. Hu, C. Song, S. Ji, K. He, X. Chen, L. Wang, X. Ma, and Q.-K. Xue, Phys. Rev. B **92**, 180507(R) (2015).

[18] W. H. Zhang, X. Liu, C. H. P. Wen, R. Peng, S. Y. Tan, B. P. Xie, T. Zhang, and D. L. Feng, Nano Lett. **16**, 1969 (2016).

[19] J. Shiogai, Y. Ito, T. Mitsuhashi, T. Nojima, and A. Tsukazaki, Nat. Phys. **12**, 42 (2016).

[20] N. Peimyoo, W. Yang, J. Shang, X. She, Y. Wang, and T. Yu, ACS Nano **8**, 11320 (2014).

[21] W. Chen, S. Chen, D. C. Qi, X. Y. Gao, and A. T. S. Wee, J. Am. Chem. Soc. **129**, 10418 (2007).

[22] J. T. Sun, Y. H. Lu, W. Chen, Y. P. Feng, and A. T. S. Wee, Phys. Rev. B **81**, 155403 (2010).

[23] T. Bathon, P. Sessi, K. A. Kokh, O. E. Tereshchenko, and M. Bode, Nano Lett. **15**, 2442 (2015).

[24] I. Carmeli, A. Lewin, E. Flekser, I. Diamant, Q. Zhang, J. Shen, M. Gozin, S. Richter, and Y. Dagan, Angew. Chem., Int. Ed. **51**, 7262 (2012).

[25] W. Chen, D. Qi, X. Gao, and A. T. S. Wee, Prog. Surf. Sci. **84**, 279 (2009).

[26] X. Liu, D. Liu, W. Zhang, J. He, L. Zhao, S. He, D. Mou, F. Li, C. Tang, Z. Li, L. Wang, Y. Peng, Y. Liu, C. Chen, L. Yu, G. Liu, X. Dong, J. Zhang, C. Chen, Z. Xu, X. Chen, X. Ma, Q. Xue, and X. J. Zhou, Nat. Commun. **5**, 5047 (2014).

[27] G. Kresse and J. Hafner, Phys. Rev. B **47**, 558 (1993).

[28] G. Kresse and J. Hafner, Phys. Rev. B **49**, 14251 (1994).

[29] G. Kresse and J. Furthmüller, Phys. Rev. B **54**, 11169 (1996).

[30] G. Kresse and D. Joubert, Phys. Rev. B **59**, 1758 (1999).

[31] J. P. Perdew, K. Burke, and M. Ernzerhof, Phys. Rev. Lett. **77**, 3865 (1996).

[32] P. E. Blöchl, O. Jepsen, and O. K. Andersen, Phys. Rev. B **49**, 16223 (1994).

[33] Y.-L. Wang, J. Ren, Can-Li Song, Y.-P. Jiang, L.-L. Wang, K. He, X. Chen, J.-F. Jia, S. Meng, E. Kaxiras, Q.-K. Xue, and X.-C. Ma, Phys. Rev. B **82**, 245420 (2010).

[34] J. Ren, S. Meng, Y.-L. Wang, X.-C. Ma, Q.-K. Xue, and E. Kaxiras, J. Chem. Phys. **134**, 194706 (2011).

[35] W. Tang, E. Sanville, and G. Henkelman, J. Phys.: Condens. Matter **21**, 084204 (2009).

[36] F.-C. Hsu, J.-Y. Luo, K.-W. Yeh, T.-K. Chen, T.-W. Huang, P. M. Wu, Y.-C. Lee, Y.-L. Huang, Y.-Y. Chu, D.-C. Yan, and M.-K. Wu, Proc. Nat. Acad. Sci. USA **105**, 14262 (2008).

[37] Y. J. Song, J. B. Hong, B. H. Min, Y. S. Kwon, K. J. Lee, M. H. Jung, and J.-S. Rhyee, J. Korean Phys. Soc. **59**, 312 (2011).